\pdfoutput=1
\documentclass[aps,showpacs,superscriptaddress,preprintnumbers,nofootinbib,notitlepage]{revtex4-1}

\usepackage[normalem]{ulem}
\usepackage{amsmath}
\usepackage{amssymb}
\usepackage{epsfig}
\usepackage{graphicx}
\usepackage{hyperref}
\usepackage{tabu}
\usepackage{boldline}
\usepackage{xspace}
\usepackage{floatrow}
\usepackage{slashed}
\usepackage{multirow}
\usepackage{diagbox}
\usepackage{siunitx}
\usepackage{subfig}
\usepackage{color}
\usepackage[table]{xcolor}
\usepackage{enumitem}
\usepackage[utf8]{inputenc}
\usepackage{colortbl}
\definecolor{nicered}{rgb}{0.7,0.1,0.1}
\definecolor{nicegreen}{rgb}{0.1,0.5,0.1}
\definecolor{red}{rgb}{1.0, 0, 0}

\usepackage{lineno}

\definecolor{LightCyan}{rgb}{0.88,1,1}
\definecolor{piggypink}{rgb}{0.99, 0.87, 0.9}
\definecolor{applegreen}{rgb}{0.55, 0.71, 0.0}
\definecolor{darkpastelgreen}{rgb}{0.01, 0.75, 0.24}
\definecolor{green-yellow}{rgb}{0.68, 1.0, 0.18}

\newcommand{\beq}{\begin{equation}}
\newcommand{\eeq}{\end{equation}}
\newcommand{\beqa}{\begin{eqnarray}}
\newcommand{\eeqa}{\end{eqnarray}}

\graphicspath{{figs/}}

\begin{document}

\title{Searching for Muonphilic Dark Sectors with Proton Beams}

\author{Claudia Rella}
\email{claudia.rella@unige.ch}
\affiliation{D\'epartement de Physique Th\'eorique, Universit\'e de Gen\`eve, CH-1211 Gen\`eve, Switzerland}

\author{Babette D\"obrich}
\affiliation{CERN, CH-1211 Gen\`eve 23, Switzerland}
\affiliation{Max-Planck-Institut f\"ur Physik (Werner-Heisenberg-Institut), 80805 M\"unchen, Germany}

\author{Tien-Tien Yu}
\affiliation{Department of Physics and Institute for Fundamental Science, University of Oregon, Eugene, Oregon 97403, USA}

\begin{abstract}
Proton beam-dump experiments are a high-intensity source of secondary muons and provide an opportunity to probe muon-specific dark sectors. 
We adopt a simplified-models framework for an exotic light scalar particle coupling predominantly or exclusively to muons.
Equipped with state-of-the-art muon simulations, we compute the sensitivity reach in the parameter space $(m_S, \, g_\mu)$ of the dark mediator, examining in detail the examples of the experiment NA62 in beam-dump mode and the proposed experiment SHiP. 
We find a significant yield of such exotics in the sub-GeV mass range. 
Our projections are competitive with those of primary muon-beam experiments and complementary to current constraints, spanning uncharted parameter space and accessing new physics potentially responsible for the $(g-2)_\mu$ anomaly.
\end{abstract}

\maketitle

\section{Introduction}\label{sec:intro}

Although the Standard Model (SM) of particle physics has provided a theoretically consistent description of all the known particles and their interactions, with the exception of gravity, there are several experimental observations that require the existence of new physics beyond the SM (BSM). Such observations include neutrino oscillations, and therefore neutrino masses, the existence of dark matter, as well as various experimental anomalies, all of which hint at a potential dark sector consisting of particles that do not interact with the known SM forces. A multitude of ways to explore the dark sector~\cite{Battaglieri:2017aum,Beacham:2019nyx,Essig:2022yzw} has been proposed and investigated. In this work we will focus on the use of beam-dump experiments as a promising probe of low-mass, weakly-coupled BSM mediators.

Searching for BSM physics in beam-dump data is not a new idea, and this idea was notably pursued in the past via neutrino experiments. However, the current lack of new-physics discoveries at the highest energies, as explored at the Large Hadron Collider at CERN, has given these old tactics a renewed interest. 
Among others, two factors push this development. Firstly, a plethora of possible BSM mediators at the MeV-GeV scale is motivated through the past and recent work of theorists (see, {\it e.g.}, Refs.~\cite{Alekhin:2015byh,Alexander:2016aln,Battaglieri:2017aum} and references therein). Secondly, several experimental anomalies suggest a connection to weakly-coupled, low-mass dark sectors. Examples of such potential signatures have been found in flavor physics\footnote{An overview of the current state of understanding of flavor anomalies has been presented in a 2021 workshop at CERN~\cite{flavor}.}, while measurements of decays of $^8$Be hint at a new $X17$ boson~\cite{PhysRevLett.116.042501}. 
Another example is the long-standing muon magnetic anomaly $(g-2)_\mu$, which describes the currently observed discrepancy between the empirical measurement and the SM prediction of the anomalous muon magnetic moment $a_\mu$. The most recent combined measurements of $a_\mu$ from the Fermilab National Accelerator Laboratory and the Brookhaven National Laboratory are in tension with theoretical predictions at the level of $4.2 \, \sigma$~\cite{PhysRevLett.126.141801}, where the experimental measurements give a larger value than the theory prediction.
Consequently, various studies have been performed to evaluate the (often competing) sensitivities of experiments that can be run in beam-dump mode~\cite{Beacham:2019nyx} to the detection of exotic long-lived, low-mass muon-specific particles.
Arguably, the most direct way to investigate the $(g-2)_\mu$ anomaly is via the experimental probe of new-physics mediators which (exclusively) couple to muons~\cite{Kinoshita:1990aj,Muong-2:2006rrc,Lindner:2016bgg}.
To this end, a primary muon beam can be used, as it has been proposed for, {\it e.g.}, NA64-$\mu$~\cite{Gninenko:2014pea,Araki:2014ona,Chen:2017awl,Chen:2018vkr,PhysRevD.105.052006} at CERN, as well as $M^3$~\cite{Kahn:2018cqs} and FNAL-$\mu$~\cite{Chen:2017awl} at Fermilab, or for a future muon collider~\cite{Cesarotti:2022ttv}.

\begin{figure}[htb!]
 \includegraphics[width=0.75\textwidth]{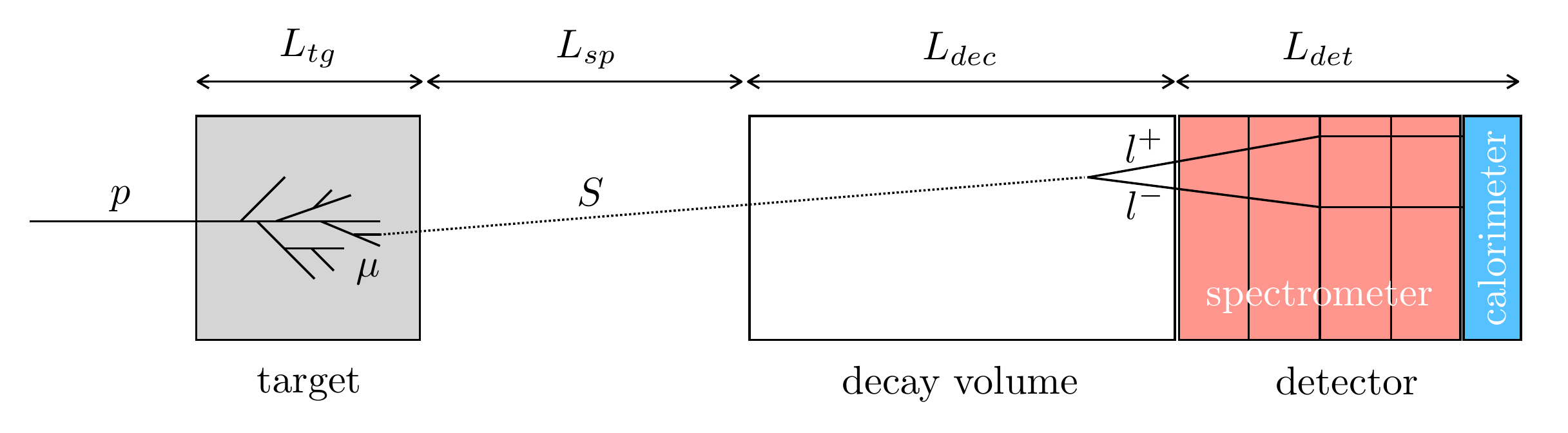}
 \caption{Schematic illustration of the framework/proposal for muonphilic scalar detection studied in this paper. Scalars emerging from a muon shower in a proton beam dump travel to a decay volume. The scalar's decay products are recorded through calorimetric and spectrometric information. Lengths are not to scale.}
 \label{fig:sketch}
\end{figure}

In this paper we aim to highlight and investigate the possibility to achieve competitive results in the detection of exotic particles radiated via bremsstrahlung of the secondary muons produced at proton beam dumps. 
The experimental setup is illustrated in Fig.~\ref{fig:sketch}.
Protons are dumped into a thick target and produce a muon shower. A secondary muon, before being possibly stopped or deflected by magnets, can emit a light scalar, which in turn travels a certain distance to the experiment's decay volume. Therein the scalar decays, either into a di-lepton final state as shown in the example in Fig.~\ref{fig:sketch} or into a pair of photons. The decay products are detected mainly through their signature in a spectrometer and a calorimeter. Details are given in Section~\ref{sec:exp_acceptance}.

Similar studies have been performed for the secondary muons produced at electron beam-dump experiments~\cite{Marsicano:2018vin}. The idea of employing secondary muons from proton beam dumps was mentioned in Ref.~\cite{Batell:2016ove} and subsequently proposed and estimated for the specific case of the SeaQuest/DarkQuest experiment in Ref.~\cite{Berlin:2018pwi}.
With this work we provide the missing sensitivity projections for two other major players that can probe new physics in this way, that is, the NA62 experiment run in beam-dump mode\footnote{A novel proposal for an experiment called ``SHADOWS", to be run in conjunction with NA62 in beam-dump mode, has recently emerged~\cite{Baldini:2021hfw}. Our studies could easily be extended to include this experiment.} and the proposed SHiP experiment. 
Crucially, for both scenarios, we aim to produce an estimate which is realistically close to what can be expected in practice.

In Section~\ref{sec:theory}, we review two simplified models of muon-specific new-physics light scalars, which can potentially explain the $(g-2)_\mu$ anomaly.
In Section~\ref{sec:cross_section}, we detail the production mechanism of the exotic scalars via muon bremsstrahlung.
In Section~\ref{sec:muons}, we describe the modelling of the secondary muon flux from the proton beam dump, which is used as input for the radiation of the exotics.
In Section~\ref{sec:exp_acceptance}, we describe the specific experimental setups of NA62 and SHiP. We include the experimental acceptance of the scalar's daughter particles, which results from a dedicated detector modelling.
In Section~\ref{sec:results}, we outline the procedure for the final sensitivity evaluation, and we show the prospective reach in the exotic scalar parameter space of the two selected proton beam-dump experiments NA62 and SHiP in the context of complementary searches. Finally, we conclude and further discuss our results in Section~\ref{sec:conclusions}.

\section{Theoretical Models of Muonphilic Dark Sectors}\label{sec:theory}

The current discrepancy between the Standard Model prediction and the experimental observation of the anomalous magnetic moment of the muon $a_\mu \equiv (g-2)_\mu/2$ is given by~\cite{PhysRevLett.126.141801}
\beq{
\Delta a_\mu \equiv a_\mu^{\rm obs}-a_\mu^{\rm SM} = (251\pm59)\times 10^{-11}\, .
}\eeq
One possible resolution of such discrepancy is obtained by introducing new-physics light particles that couple predominantly to muons. Note that particles with parity-odd couplings to muons, {\it i.e.}, pseudo-scalar and axial-vector couplings, give a negative contribution to the predicted value of $a_\mu$, which further exacerbates the discrepancy with the experimental measurement. Thus, we focus our attention on scalar and vector particles, which have parity-even couplings.
A generic light scalar $S$ or vector $V$ can couple to muons through the operators
\beq
g_{\mu, \, S} S\bar\mu\mu~~{\rm (scalar)} \, ,~~~~~g_{\mu, \, V} V_\alpha\bar\mu\gamma^\alpha\mu~~\rm{(vector)} \, ,
\eeq
where $g_{\mu, \, S}, \, g_{\mu, \, V}$ are the couplings between $S, \, V$ and muons, respectively.
These operators contribute at leading-order to the muon anomalous magnetic moment as~\cite{Leveille:1977rc}
\beq{
\Delta a_\mu^S=\frac{g_{\mu, \, S}^2}{8\pi^2}\int_0^1 dz\frac{m_\mu^2(1-z)^2(1+z)}{m_\mu^2(1-z)^2+m_S^2z}\simeq 2.3\times 10^{-9}\left(\frac{g_{\mu, \, S}}{3.5\times10^{-4}}\right)^2~~~(m_S\ll m_\mu) \, ,
}\eeq
and
\beq{
\Delta a_\mu^V=\frac{g_{\mu, \, V}^2}{4\pi^2}\int_0^1 dz\frac{m_\mu^2z(1-z)^2}{m_\mu^2(1-z)^2+m_V^2z}\simeq 2.1\times 10^{-9}\left(\frac{g_{\mu, \, V}}{5.0\times10^{-4}}\right)^2~~~(m_V\ll m_\mu) \, .
}\eeq

The addition of a new $U(1)$ gauge group is a natural extension of the SM and results in a new vector boson. The most straightforward way to obtain this is through the ``kinetic-mixing" portal that mixes the $U(1)_Y$ of SM hypercharge with a $U(1)_D$. Although the associated vector boson, known as the dark photon, does provide a positive contribution to $a_\mu$, the parameters needed to alleviate the $(g-2)_\mu$ anomaly are inconsistent with direct searches for the dark photon (see, {\it e.g.}, Ref.~\cite{NA482:2015wmo}). 
Another proposed vector model extends the SM with the anomaly-free $U(1)_{L_\mu-L_\tau}$ with corresponding vector boson $Z^\prime$. This new vector boson couples to $\mu, \, \tau$ leptons as well as their corresponding neutrino flavors $\nu_\mu, \, \nu_\tau$. 
Although the $L_\mu-L_\tau$ vector boson~\cite{He:1990pn, Foot:1990mn, He:1991qd} is an attractive candidate, its primary decays are through neutrinos. 
Due to the difficulty of reconstructing the so-called ``open decays'' involving neutrinos in a beam-dump setup, we do not evaluate this scenario in what follows.
However, our analysis can be straightforwardly extended to include this possibility in case these invisible decay modes can be studied experimentally.
For the remainder of this work, we will only consider scalar candidates.

As discussed above, a new light scalar particle coupling to muons will increase the value of $a_\mu$ and thus help alleviate the tension in the muon magnetic anomaly $(g-2)_\mu$. At low-energies, the relevant part of the Lagrangian is given by
\beq{ \label{eq:simplified_model}
{\cal L}\supset \frac{1}{2}(\partial_\alpha S)^2-\frac{1}{2}m_S^2 S^2-\sum_{\ell=e,\mu,\tau}g_\ell S\bar\ell \ell \, ,
}\eeq
where $g_\ell$ is the coupling between the scalar $S$ and the leptons $\ell=e,\, \mu, \, \tau$, and $m_S$ is the mass of $S$. Although the effective Lagrangian in Eq.~\eqref{eq:simplified_model} does not respect the $SU(2) \times U(1)$ gauge symmetry of the SM, it can be generalised to the effective dimension-5 operator ${\cal O}_5=\frac{1}{\Lambda}(\bar L E) H S$, where $H$ is the SM Higgs doublet, and $L, \, E$ are the lepton doublets and singlets, respectively. The operator ${\cal O}_5$ does respect the $SU(2) \times U(1)$ gauge symmetry, and it can in turn be UV completed with, {\it e.g.}, vector-like fermions~\cite{Fox:2011qd} or multiple Higgs states~\cite{Chen:2015vqy,Batell:2016ove,PhysRevD.98.055026}. For the purpose of this work we will remain agnostic about the UV-completion of the simplified model in Eq.~\eqref{eq:simplified_model}, and we will regard the couplings $g_{\ell}$, $\ell=e,\, \mu,\, \tau$, of the exotic particle $S$ to leptons as free parameters of the theory.
%
From the model-building point of view, there are substantial motivations to consider $g_\ell \propto m_\ell$, where $m_\ell$ is the mass of the lepton. 
In particular, introducing new lepton-specific scalar mediators below the electroweak scale can lead to large flavor-changing neutral currents (FCNCs), which are strongly constrained by null searches for $\mu-e$ conversion, $\mu\to 3e$, and $\mu\to e\gamma$~\cite{SINDRUMmu2e,MEG:2016leq,ParticleDataGroup:2020ssz}. Requiring the mass proportionality $g_\ell\propto m_\ell$, a condition known as minimal flavor violation (MFV)~\cite{Cirigliano:2005ck}, avoids FCNCs. We will denote this scenario as the ``leptophilic model".
Alternatively, we can further impose $g_e=g_\tau=0$, such that the scalar exclusively couples to muons. In this case, suppression of the FCNCs can be ensured by, {\it e.g.}, considering radiatively generated FCNCs~\cite{PhysRevD.98.055026}, which are shown to be suppressed by small Yukawa couplings. We will denote this scenario as the ``muonphilic model". This muon-only framework is particularly challenging to probe but not unfeasible, as we will see, for collider and electron or proton beam-dump searches.

In both the leptophilic and the muonphilic models, the width of the scalar decay into leptons $S \rightarrow \ell^+ \ell^-$ is given by
\beq{
\Gamma_{\ell^+\ell^-}=\frac{m_S}{8\pi}g_\ell^2\left(1-\frac{4m_\ell^2}{m_S^2}\right)^{3/2} \, ,
}\eeq
while the decay into photons $S \rightarrow \gamma \gamma$ arises from the effective scalar coupling to vector boson pairs at one-loop level, and its decay width is given by 
\beq{ \label{eq:sgammagamma}
\Gamma_{\gamma\gamma}= \frac{\alpha^2m_S^3}{64\pi^3}\left|\sum_{\ell=e,\mu,\tau}\frac{g_\ell}{m_\ell}x_\ell[1+(1-x_\ell)f(x_\ell)]\right|^2,
}\eeq
where $\alpha=1/137$ is the fine-structure constant, $x_\ell=4m_\ell^2/m_S^2$, and 
\beq{
f(x)=\begin{cases}
\arcsin^2({x^{-1/2}}) \, , & ~~x > 1\\
-\frac{1}{4}\left[\ln\left(\frac{1+\sqrt{1-x}}{1-\sqrt{1-x}}\right)-i\pi\right]^2 \, , & ~~x\leq 1
\end{cases} \, .
}\eeq
Note that, in the muonphilic scenario, in which $g_e=g_\tau=0$, the decay into photons results from the muon loop contribution only, which is
\beq{
\Gamma_{\gamma\gamma}^{\mu-{\rm loop}}=\frac{\alpha^2m_S^3}{64\pi^3}\left|\frac{g_\mu}{m_\mu}x_\mu[1+(1-x_\mu)f(x_\mu)]\right|^2 \, .
}\eeq
We denote by $\Gamma_S$ the total decay width of the scalar. For $2m_e < m_S < 2m_\mu$, the widths of the decay channels $S \rightarrow \gamma \gamma$ and $S \rightarrow e^+e^-$ sum to give
\beq{
\Gamma_S=\begin{cases}
\Gamma_{e^+e^-}+\Gamma_{\gamma\gamma}\simeq \Gamma_{e^+e^-} \, , & ~~g_\ell\propto m_\ell\\
\Gamma_{\gamma\gamma}^{\mu-{\rm loop}} \, , & ~~g_\mu\propto m_\mu, \, g_e=g_\tau=0
\end{cases}\, ,
}\eeq
while, for $2m_\mu < m_S < 2m_\tau$, the scalar decay into muons $S \rightarrow \mu^+\mu^-$ is dominant in both models, that is  
\beq{
\Gamma_S=\begin{cases}
\Gamma_{\mu^+\mu^-}+\Gamma_{e^+e^-}+\Gamma_{\gamma\gamma}\simeq \Gamma_{\mu^+\mu^-} \, , & ~~g_\ell\propto m_\ell\\
\Gamma_{\mu^+\mu^-}+\Gamma_{\gamma\gamma}^{\mu-{\rm loop}}\simeq \Gamma_{\mu^+\mu^-} \, , & ~~g_\mu\propto m_\mu, \, g_e=g_\tau=0
\end{cases}\, .
}\eeq
In what follows, we use the exact formulas for $\Gamma_S$.
The corresponding decay length of the scalar is given by
\beq{
L_S=\frac{E_S}{m_S}\frac{\beta_S}{\Gamma_S} \, ,
}\eeq
where $E_S$ is the energy of the scalar and $\beta_S=\sqrt{1-m_S^2/E_S^2}$ is its boost factor. 
As an example, for a reference value of $E_S=30$ GeV, we have
\begin{align}
L_S&\simeq60.7{~\rm m}\times\left(\frac{1.0\times10^{-4}}{g_\mu}\right)^2\times \left(\frac{100~{\rm MeV}}{m_S}\right)^2~~~\text{leptophilic ($g_\ell\propto m_\ell$)}\, , \\
L_S&\simeq61.2{~\rm m}\times\left(\frac{9.0\times10^{-4}}{g_\mu}\right)^2\times \left(\frac{100~{\rm MeV}}{m_S}\right)^2~~~\text{muonphilic ($g_\mu\propto m_\mu,  \, g_e=g_\tau=0$)}\, , 
\end{align}
where the parameters are chosen to give decay lengths comparable to the minimum distance that the scalar must travel to reach the decay volume of SHiP (NA62) at 53.2 (79.0) m.

\section{Scalar Production Cross-Section}\label{sec:cross_section}

\begin{figure}[htb!]
 \includegraphics[width=0.5\textwidth]{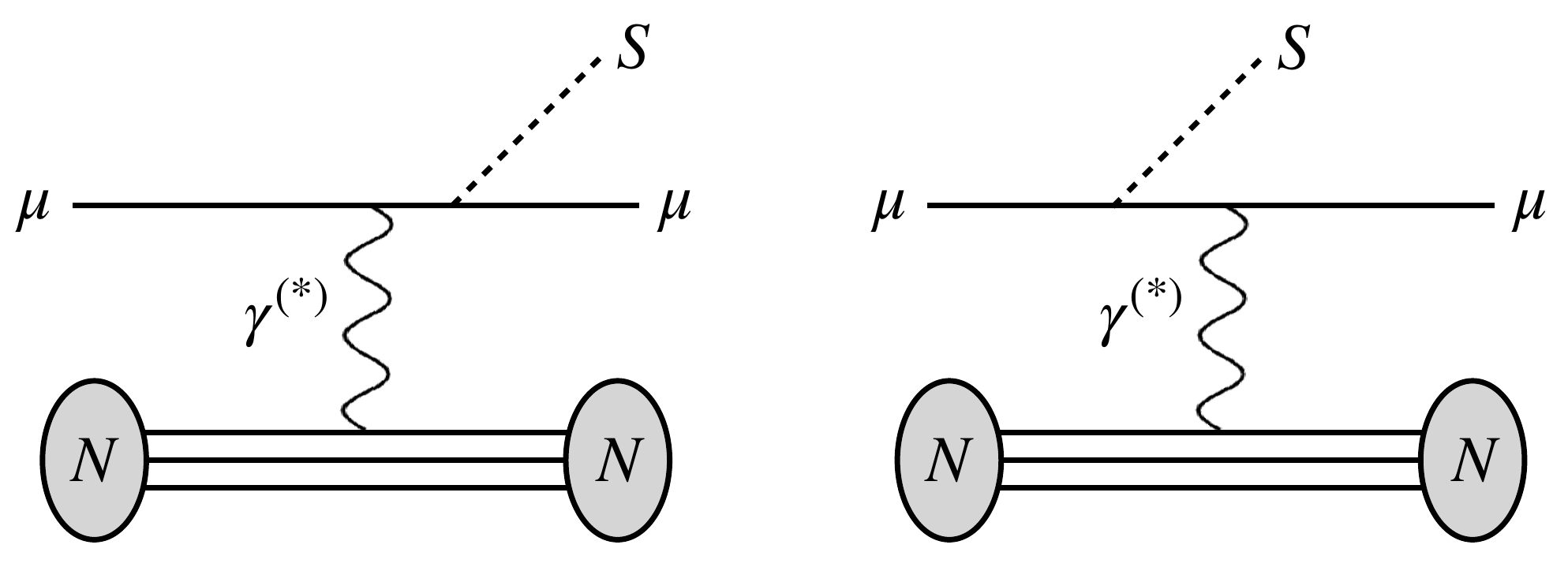}
 \caption{Feynman diagrams corresponding to the radiative production of the scalar $S$ from muon bremsstrahlung.}
 \label{fig:scalar_brem}
\end{figure}

The dominant production mechanism for the new light scalar state $S$ at beam dumps is the $\mu+N\to \mu+N+S$ bremsstrahlung process shown in Fig.~\ref{fig:scalar_brem}, where the incident muon\footnote{Here, $\mu$ refers to both $\mu^+$ and $\mu^-$.} $\mu$ exchanges a virtual photon $\gamma^{(*)}$ with the target nucleon $N$ and radiates the exotic scalar $S$~\cite{PhysRevD.80.075018}.

When the beam energy is much higher than the masses of both the beam and the radiation, in our case the muon and the scalar, respectively, the signal production cross-section can be estimated by the Weizs\"acker-Williams (WW) approximation~\cite{Weizsacker:1934, Williams:1934}.
In this scenario of highly-boosted incoming muons, the intermediate virtual photons produced by the muons scattering off the target nuclei are nearly on-shell and can be approximated by real photons. Thus, the phase space integration of the full $2 \rightarrow 3$ scattering process $\mu(p) + N(P_i) \to \mu(p') + N(P_f) + S(k)$ can be estimated via the simpler $2 \rightarrow 2$ process $\mu(p) + \gamma(q) \to \mu(p') + S(k)$, evaluated at minimum virtuality $t_{\rm min} \equiv q^2_{\rm min}$ and weighted by the effective photon flux $\chi$. 
Applying the WW approximation scheme under the assumptions $E_\mu \gg m_\mu, m_S$, the differential production cross-section in the lab frame is given by
\beq{ \label{eq:diffsigma}
\frac{d\sigma}{dx}(\mu+N\to \mu+N+S)\simeq\frac{g_{\mu}^2\alpha^2}{12\pi}\chi\beta_S\beta_\mu\frac{x^3[m_\mu^2(3x^2-4x+4)+2m_S^2(1-x)]}{[m_S^2(1-x)+m_\mu^2x^2]^2} \, ,
}\eeq
where $x=E_{S}/E_\mu$ is the fraction of the incoming muon energy $E_\mu$ taken by the emission of an exotic scalar of energy $E_S$. The boost factors are $\beta_{S}=\sqrt{1-m_{S}^2/(xE_\mu)^2}$ and $\beta_\mu=\sqrt{1-m_\mu^2/E_\mu^2}$.
The effective photon flux $\chi$ is defined as
\beq{ \label{eq:chi_flux}
\chi=\int_{t_{\rm min}}^{t_{\rm max}}dt\frac{t-t_{\rm min}}{t^2}G_2(t) \, ,
}\eeq 
where the virtuality $t$ is the momentum transfer squared and $G_2(t)$ is the combined atomic and nuclear electric form factor of the target atom, which includes both elastic and inelastic contributions~\cite{PhysRevD.80.075018}.
More precisely, we have $G_2(t)=G_2^{\rm el}(t)+G_2^{\rm in}(t)$, where the elastic and inelastic components are given by
\begin{align}
G_2^{\rm el}(t)&=\left(\frac{a^2 t}{1+a^2 t}\right)^2Z^2\left(\frac{1}{1+t/d}\right)^2 \, , \\
G_2^{\rm in}(t)&=\left(\frac{a^{\prime 2}t}{1+a^{\prime 2}t}\right)^2 Z\left(\frac{1+t \, (\mu_p^2-1)/(4m_p^2)}{(1+t/(0.71~\rm{GeV}^2))^4}\right)^2 \, ~~~~\text{($t/m_p^2 \ll 1$)} \, ,
\end{align}
where $a=111 \, Z^{-1/3}/m_e$ in the Thomas-Fermi model, $d=0.164~{\rm GeV}^2 A^{-2/3}$, $a^\prime=773 \,  Z^{-2/3}/m_e$, and $\mu_p=2.79$. $m_e, \, m_p$ are the electron and proton masses, and $A, \, Z$ are the atomic mass and number of the target material.

\begin{figure}[htb!]
 \includegraphics[width=0.6\textwidth]{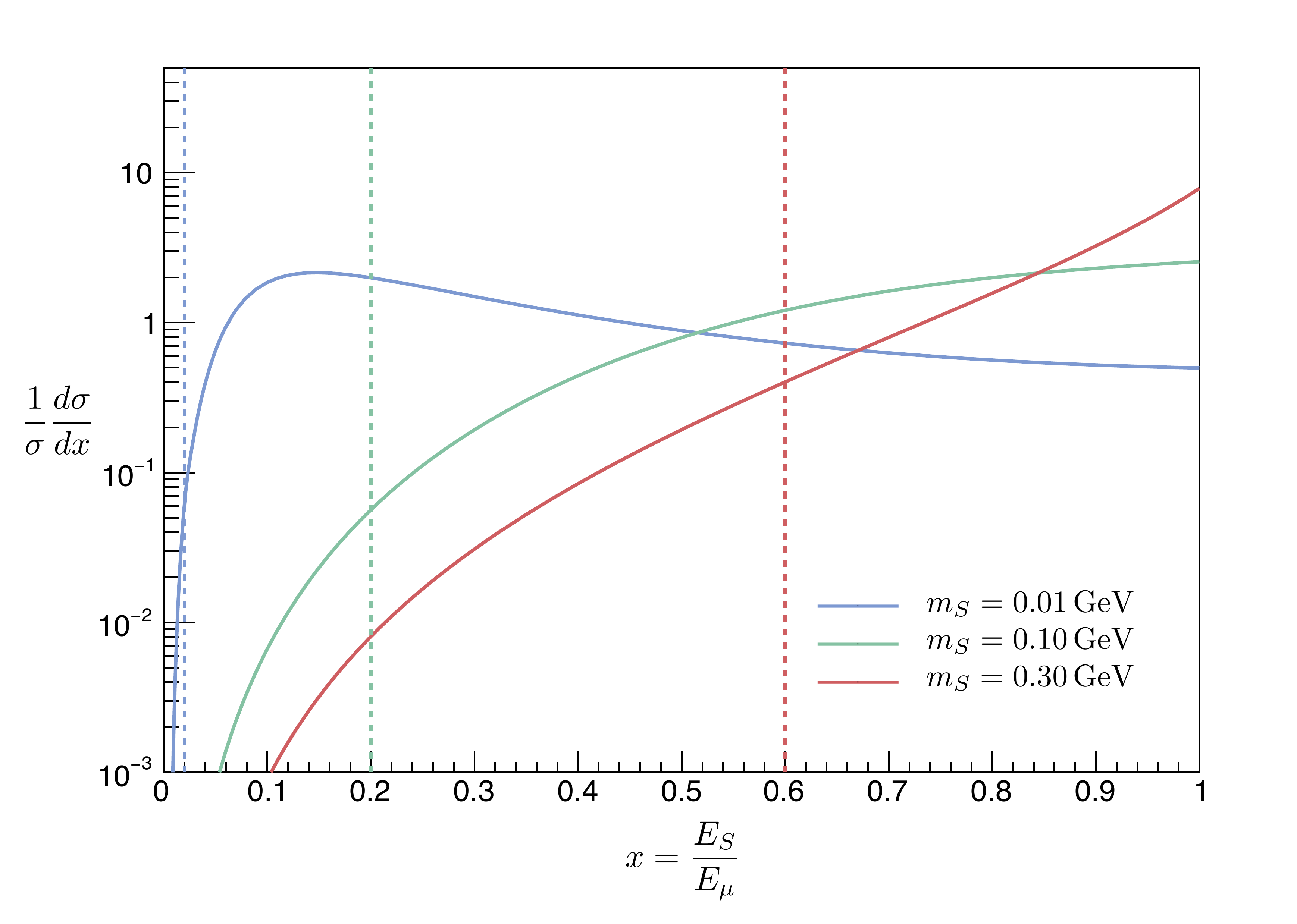}
 \caption{The solid curves are the distributions of the fraction $x$ of the incoming muon energy taken by the emission of the exotic scalar $S$ for various scalar masses. The dashed vertical lines correspond to the values of $x$ where the Lorentz factor $\gamma_S = 100$ for each $m_S$ and a reference choice of $E_\mu = 50 \, \text{GeV}$. We show $m_S = 0.01 \; \text{GeV}$ in blue, $m_S = 0.10 \; \text{GeV}$ in green, and $m_S = 0.30 \; \text{GeV}$ in red.}
 \label{fig:sigma_ms}
\end{figure}

The normalised distribution of Eq.~\eqref{eq:diffsigma}, which reproduces the expected energy spectrum of the new scalar mediator, changes significantly with the scalar mass. As shown in Fig.~\ref{fig:sigma_ms}, the peak of the distribution shifts from low to high values of $x$ as $m_S$ increases. Moreover, for each choice of $m_S$ and for $E_\mu \gg m_S$, the distribution lies dominantly in the region of high Lorentz factor $\gamma_S = (1-\beta_S^2)^{-1/2} \gg 1$, which corresponds to the right of the matching dashed vertical line in Fig.~\ref{fig:sigma_ms}.
Thus, for sufficiently high values of the muon energy, the emission of the exotic scalar particle $S$ happens primarily in the highly relativistic regime.

A refinement of the WW approximation follows from the observation that, in the domain of highly relativistic particles, the radiation is dominantly collinear with the beam.
Then, the phase space integration in Eq.~\eqref{eq:chi_flux} can be further facilitated, yielding the Improved Weizs\"acker-Williams (IWW) approximation~\cite{Liu:2016mqv, PhysRevD.8.3109}, which is implemented in the simplification of the integration limits in the virtuality as
\beq{ \label{eq:chi_flux2}
\chi \simeq \int_{m_S^4/(4E_\mu^2)}^{m_S^2+m_\mu^2}dt\frac{t-m_S^4/(4E_\mu^2)}{t^2}G_2(t) \, .
}\eeq 
We note that the WW approximated expression for the scalar production differential cross-section in Eq.~\eqref{eq:diffsigma} and the improved effective photon flux integration in Eq.~\eqref{eq:chi_flux2} are only valid for muon energies well above the muon and scalar masses, \textit{i.e.} $E_\mu \gg m_\mu, m_S$.
To check the applicability of the IWW approximation to our context, and to roughly estimate its accuracy, we compute the normalised distribution in $x$ of the scalar production differential cross-section $\frac{1}{\sigma} \frac{d\sigma}{dx}$ using {\tt MadGraph5}~\cite{Alwall:2014hca} (MG) for selected benchmark points in $m_S$ and $E_\mu$ at fixed coupling $g_\mu$.
We then compare the MG results with the corresponding distributions computed using the IWW approximation in Eqs.~\eqref{eq:diffsigma},~\eqref{eq:chi_flux2}. 
We find a discrepancy between the IWW and the MG computed cross-section distributions, quantified as their relative difference with sign, which is $\lesssim 20\%$ when $E_\mu \gtrsim {\rm max}(300 \, m_S, \, m_\mu)$, and a larger discrepancy for smaller values of the muon energy\footnote{Similar conclusions are found in Ref.~\cite{Liu:2016mqv}, where a systematic comparison of the exact tree-level calculation compared to the WW and IWW approximations is performed for scalar production at electron beam dumps. An analogous study for vectors is presented in Ref.~\cite{PhysRevD.104.076012}.}.
We additionally verify that, for such values of $E_\mu$, the assumption of highly relativistic scalars holds at the corresponding values of $m_S$.
Therefore, we adopt an energy cut on the incoming muons spectrum at $E_{\mu}^{\rm cut}={\rm max}(300 \, m_S, \, m_\mu)$, which ensures that we are in the regime of validity of the IWW approximation and that our estimates are conservative. 
We emphasise here that the purpose of this work is to present the potential of using the secondary muons produced in proton beam-dump experiments to probe new light muon-specific scalars. We leave a more detailed analysis, which should make use of an exact cross-section calculation, for future work.

\section{Modelling the Muons from the Beam Dump}\label{sec:muons}

At the heart of the sensitivity projections described in Section~\ref{sec:results} lies a thorough modelling of the original muon flux from the proton beam dump with sufficiently high statistics, to which different approaches have been explored~\cite{SHiP:2019gcl,Ghinescu:2021sjm}.
The dominant muon flux comes from decaying mesons (mostly pions and kaons) produced in the proton collisions with the dump material.
The challenge to model this flux is manifold and arises from the following requirements:
\begin{enumerate}
    \item A detailed understanding of the physical processes that eventually yield the muons;
    \item An efficient Monte Carlo (MC) simulation of those physical processes which does not suffer from statistical fluctuations.
\end{enumerate}

The muon flux used in this study is produced via {\tt GEANT4}~\cite{GEANT4:2002zbu} and the relevant code is publicly available in Ref.~\cite{stealex_2022_6352043}.
The simulated meson spectrum can be cross-checked against the secondary mesons created in proton-nucleon collisions, as modelled in Ref.~\cite{VanDijk:2641500} from measurements using thin targets. We are most interested in the shape of the muon spectrum shortly after the creation of the muons, which closely tracks the shape of the meson spectrum. Note that validating the actual muon spectrum in the downstream part of the experiment is a much more challenging task as the experiment's magnets modify the shape of the muon spectrum away from the mesons. Fortunately, we do not need to undertake such a validation of the downstream spectrum as our study does not require it. 
After validation, the simulated muons undergo a biasing mechanism, as described in Ref.~\cite{Ghinescu:2021sjm}, in order to efficiently reproduce a muon sample up to the highest muon energies.
This biasing process boosts the muon statistics by more than three orders of magnitude without increasing computing resources and without altering the physics of a ``vanilla" simulation\footnote{In previous studies~\cite{Rosenthal:ICAP2018-SUPAG05}, ``vanilla" muon simulations were parametrised and then re-sampled to gain statistics.} (see Ref.~\cite{Ghinescu:2021sjm} for details).
We note that the simulated experiment specifically mimics the features of the NA62 dump, which has a different geometry and composition than the SHiP dump.
However, as shown in Fig.~2 of Ref.~\cite{Ghinescu:2021sjm}, the relevant muons typically emerge at depths of about $50~\text{cm}$. Thus, the actual length of the dump does not matter here as long as it is at least of a few meters. Moreover, the cross-section for meson production with larger target nuclei can be obtained with an appropriate scaling of the target material as $A^{2/3}$, where $A$ is the atomic mass number\footnote{As described in Section~\ref{sec:exp_acceptance}, the NA62 target material is Copper (atomic mass of about $59.19~\text{GeV}$), while the SHiP target material is Molybdenum (atomic mass of about $89.37~\text{GeV}$).}. This implies that using the NA62-simulated input data will result in a slight underestimation of the reach of SHiP.

\begin{figure}[htb!]
 \centering
 \subfloat[NA62 Experiment]{\includegraphics[width=0.49\textwidth]{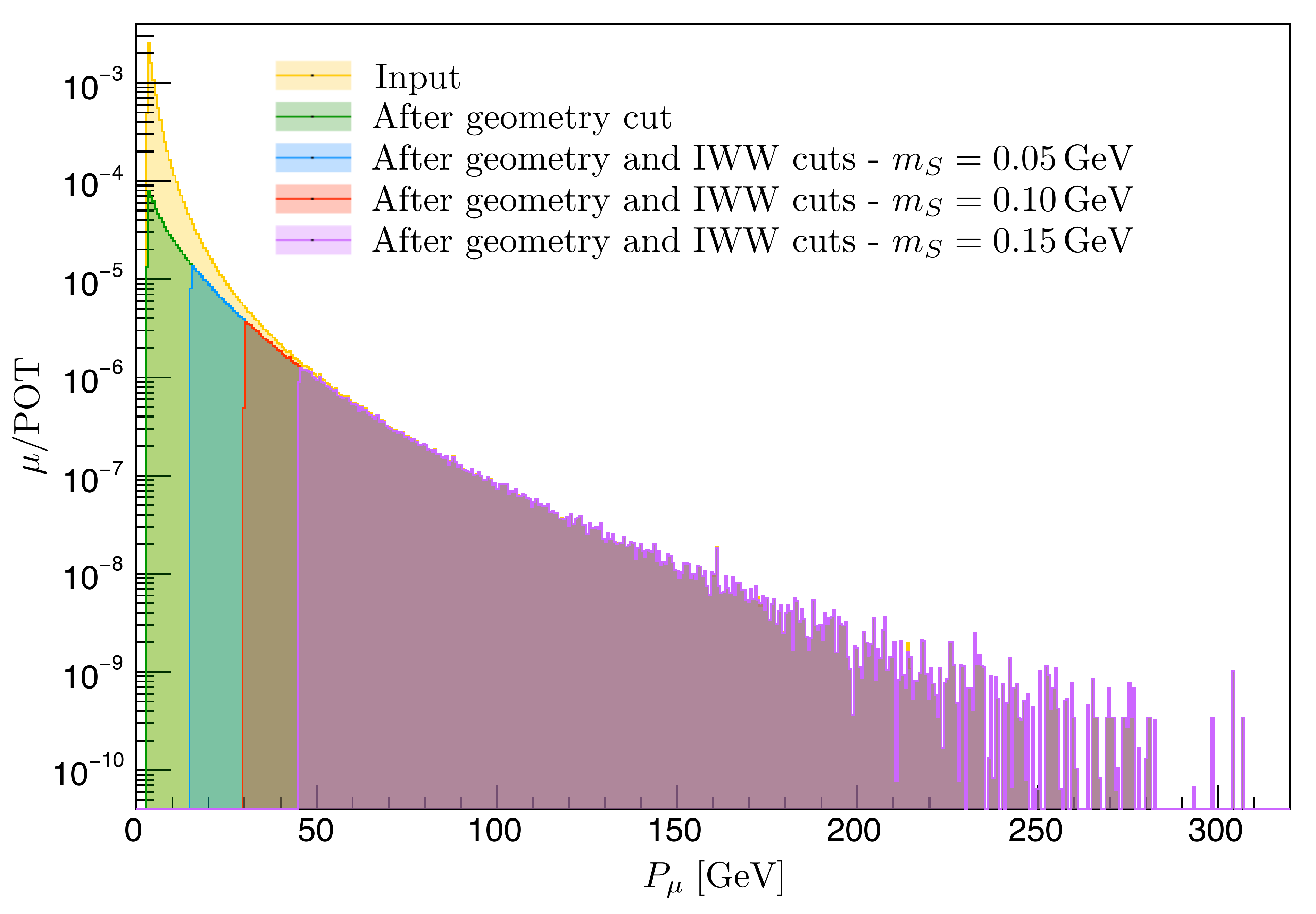}} 
 \quad 
 \subfloat[SHiP Experiment]{\includegraphics[width=0.49\textwidth]{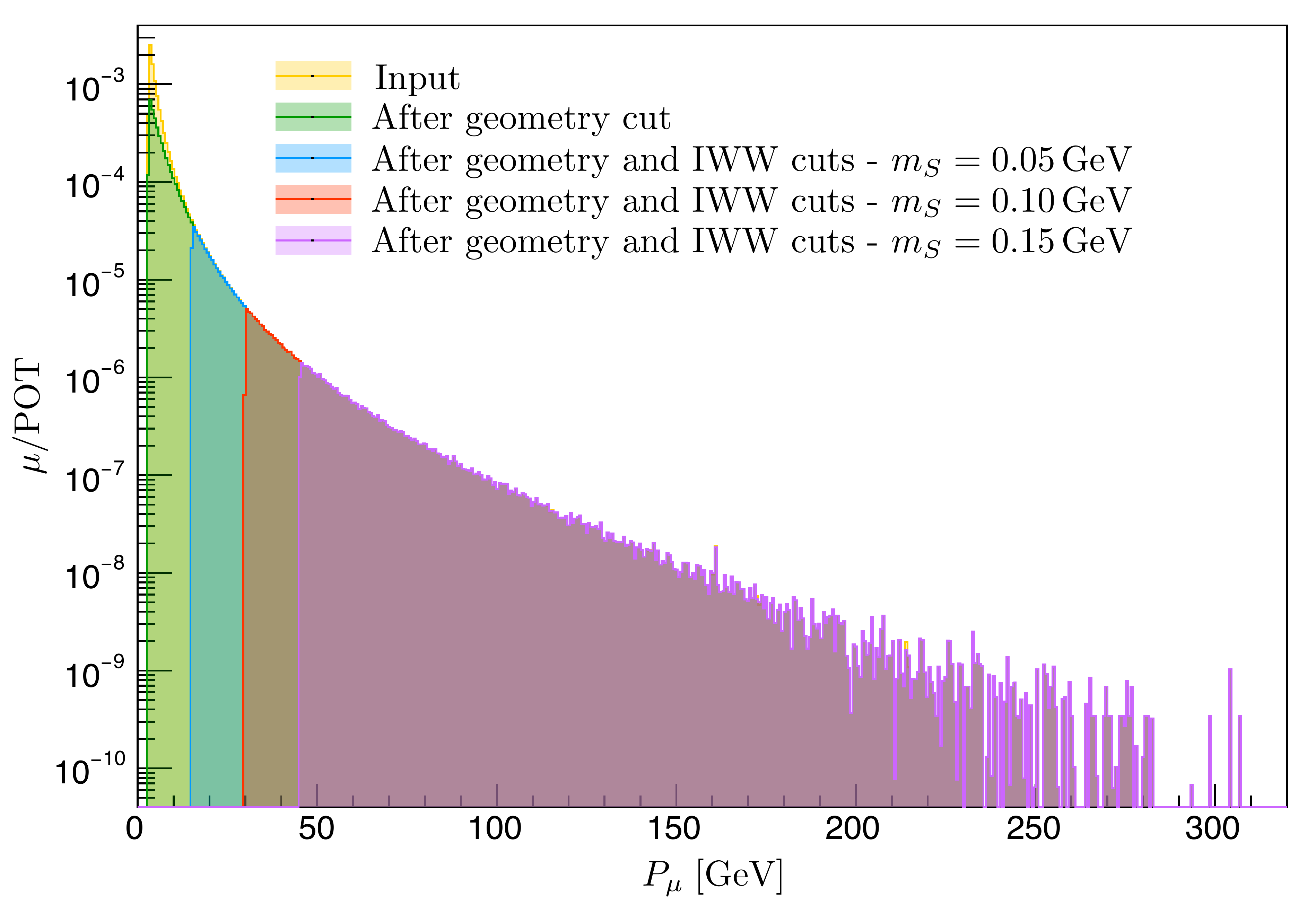}}
 \caption{Distributions of the total momentum $P_\mu$ of the secondary muons for the experiments (a) NA62 and (b) SHiP. We show the input spectra in yellow, the spectra after geometrical acceptance in green, and the spectra after geometrical acceptance and fixed-mass IWW energy cuts in blue ($m_S= 0.05 \, \text{GeV}$), red ($m_S= 0.10 \, \text{GeV}$),  and violet ($m_S= 0.15 \, \text{GeV}$).}
\label{fig:muonsmomentum}
\end{figure}

The original and effective distributions of the muons' total momentum are shown in Fig.~\ref{fig:muonsmomentum} for the experiments NA62 (left) and SHiP (right). 
The input distribution, in yellow, is obtained from $400 \, \text{GeV}$ protons at production point via MC simulation, following the techniques presented in Ref.~\cite{Ghinescu:2021sjm}, as described above.
The distributions in green are obtained from the input spectrum by imposing the geometrical acceptance of the incoming muons and the radiated scalars, as clarified in Section~\ref{sec:exp_acceptance}. 
Then, the additional IWW condition $E_\mu \gtrsim {\rm max}(300 \, m_S, \, m_\mu)$, which is explained in Section~\ref{sec:cross_section}, is applied for benchmark values of the scalar mass in order to produce the effective distributions in blue for $m_S = 0.05 \, \text{GeV}$, red for $m_S = 0.10 \, \text{GeV}$, and violet for $m_S = 0.15 \, \text{GeV}$.

We observe that both the experimental geometry cut and the IWW constraint on the muon energy primarily act on the low-momentum part of the spectrum, where the original peak occurs, and thus progressively shift the mean of the distribution to the right.
The total percentage loss of events from the yellow to the blue distributions is about $97 \, \%$ for NA62 and about $95 \, \%$ for SHiP. 
Moreover, the relatively small overall geometrical acceptance of NA62 dominates over the IWW energy cut, while the SHiP spectrum is predominantly shrunk by the IWW since its larger transversal extent allows to keep a bigger portion of the low-energy muons.
Despite the loss of events induced by the constraints under consideration, we observe that, for both experiments, for a minimum scalar mass of about $50 \, \text{MeV}$, the average effective muon flux after geometrical acceptance and energy cuts is of the order of $10^{-5} \, \mu / \text{POT}$. Then, the large number of primary protons which can be typically dumped\footnote{As described in Section~\ref{sec:exp_acceptance}, we consider $10^{19} \, \text{POT}$ for NA62 and $2 \times 10^{20} \, \text{POT}$ for SHiP.} suggests how proton beam-dump experiments might indeed be competitive signal production sources in the search for exotic muon-specific scalars compared to ``direct" muon-beam experiments such as NA64-$\mu$~\cite{Gninenko:2014pea}, as well as the proposed $M^3$~\cite{Kahn:2018cqs} and FNAL-$\mu$~\cite{Chen:2017awl} at Fermilab.

\begin{figure}[htb!]
 \centering
 \subfloat[NA62 Experiment]{\includegraphics[width=0.49\textwidth]{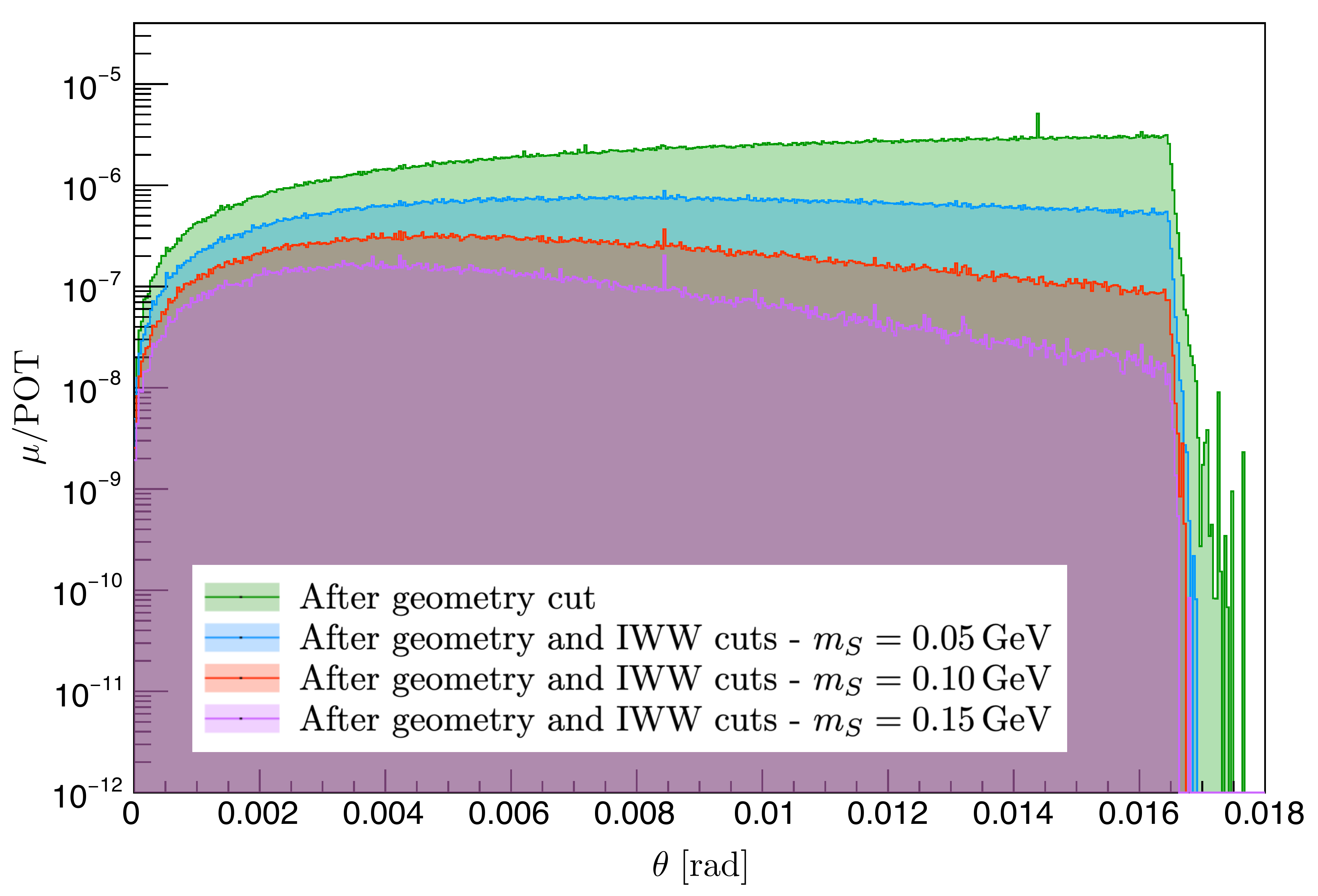}}
 \quad 
 \subfloat[SHiP Experiment]{\includegraphics[width=0.49\textwidth]{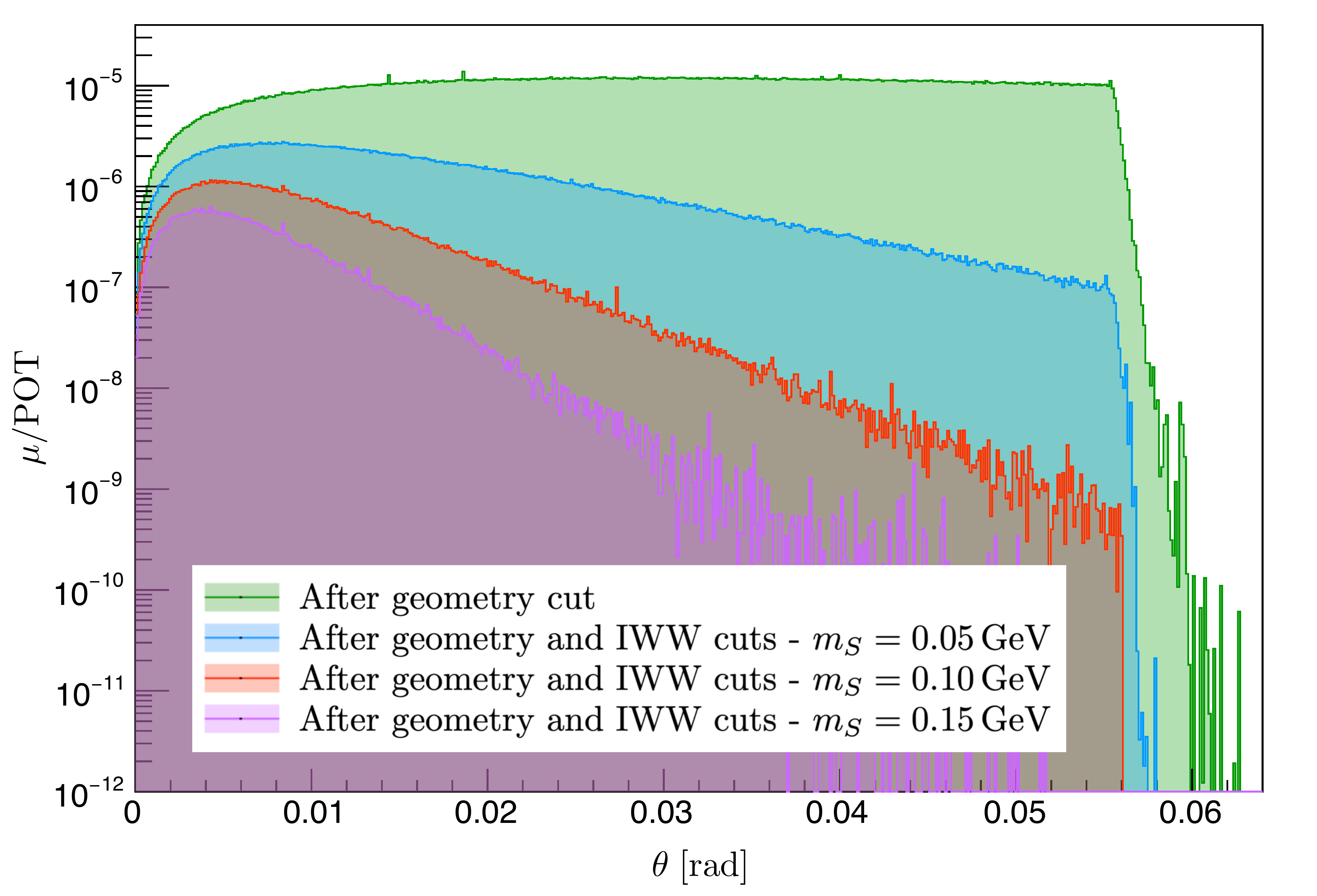}}
 \caption{Angular distributions of the secondary muons for the experiments (a) NA62 and (b) SHiP. Here, $\theta$ is the angle with respect to the proton beam axis. We show the spectra after geometrical acceptance in green, and the spectra after geometrical acceptance and fixed-mass IWW energy cuts in blue ($m_S= 0.05 \, \text{GeV}$),  red ($m_S= 0.10 \, \text{GeV}$), and violet ($m_S= 0.15 \, \text{GeV}$).}
\label{fig:muonsangle}
\end{figure}

The effective angular distributions of the secondary muons are shown in Fig.~\ref{fig:muonsangle} for the experiments NA62 (left) and SHiP (right). A schematic representation of the geometrical meaning of the angle $\theta$ of the muons' trajectories with respect to the proton beam axis is shown in Fig.~\ref{fig:geom_sketch} in Section~\ref{sec:results}. The effective distributions in green, blue, red, and violet are obtained from the original spectrum applying the same requirements detailed above for Fig.~\ref{fig:muonsmomentum}. We do not display the input angular distribution, which peaks at about $0.04 \, \text{rad}$ and extends up to about $0.9 \, \text{rad}$, for clarity of the plots.

\begin{figure}[htb!]
 \centering
 \subfloat[NA62 Experiment]{\includegraphics[width=0.49\textwidth]{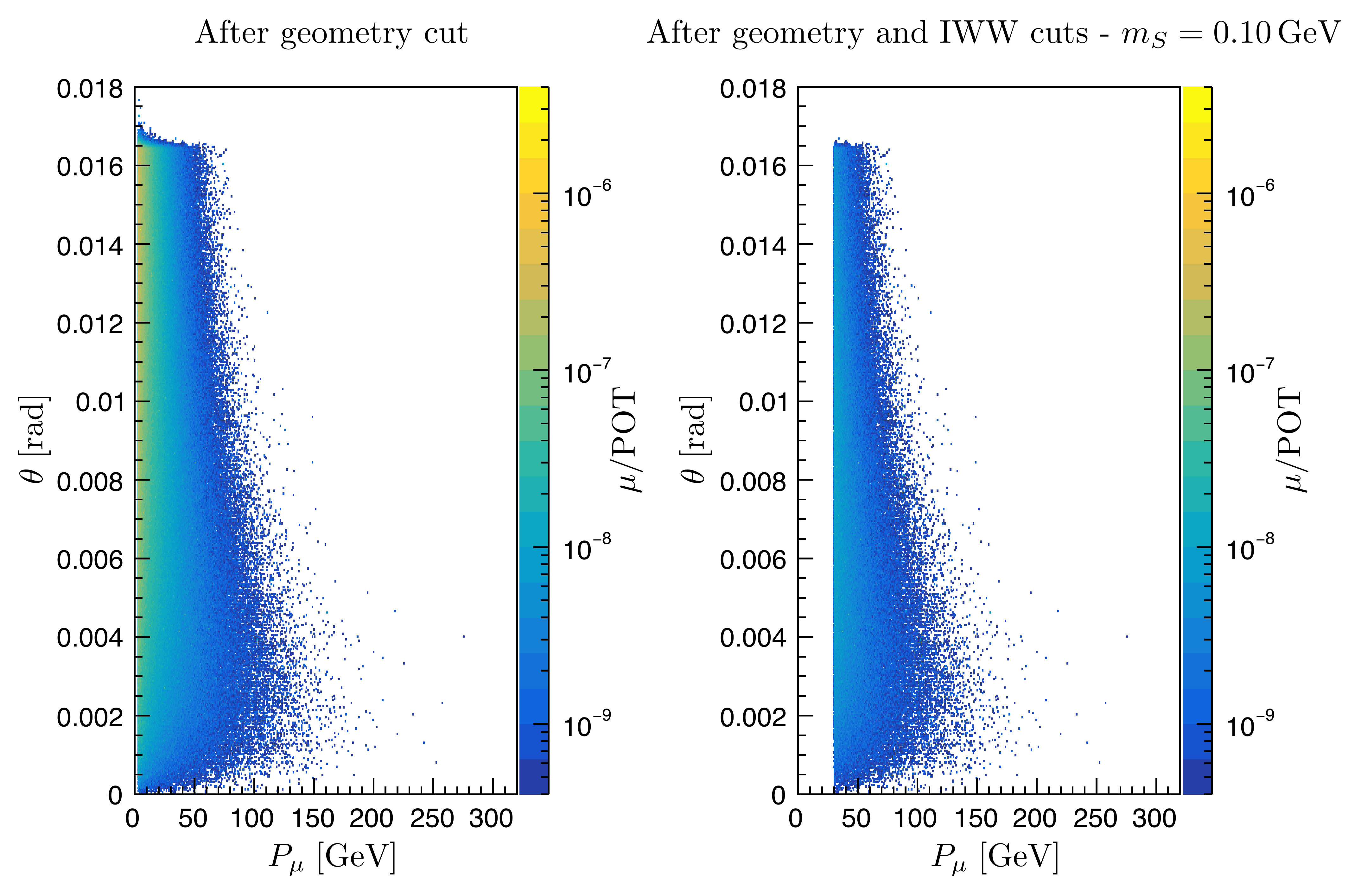}}\quad
 \subfloat[SHiP Experiment]{\includegraphics[width=0.49\textwidth]{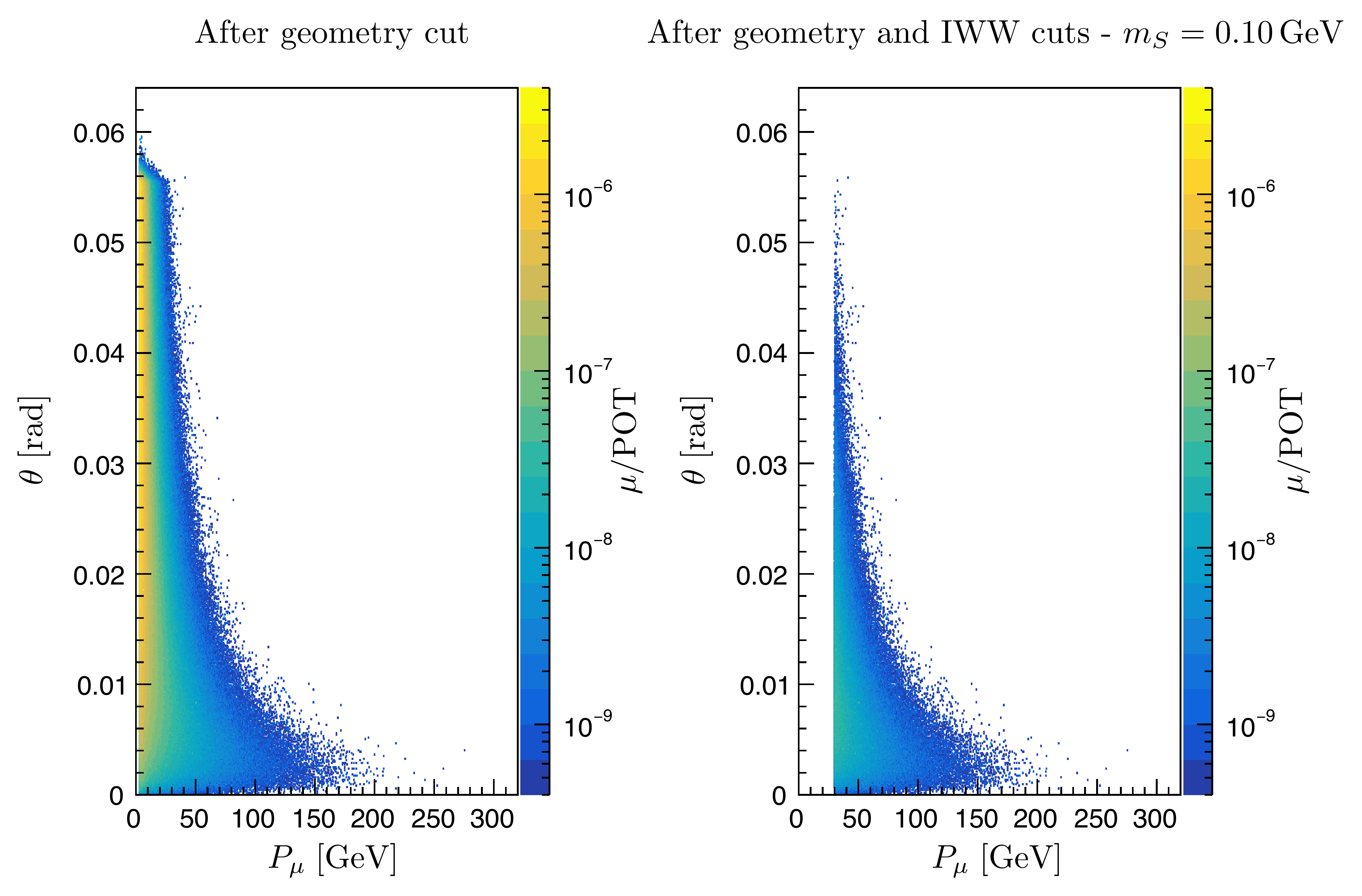}}
 \caption{Effective distributions of the secondary muons for the experiments (a) NA62 and (b) SHiP in the plane $(P_{\mu}, \, \theta)$. For each experiment, we show the spectrum after geometrical acceptance (left) and the spectrum after geometrical acceptance and IWW energy constraint for $m_S= 0.10 \, \text{GeV}$ (right). Increasing or decreasing the value of $m_S$ will shift the vertical cut due to the IWW towards higher or lower momenta, respectively.}
\label{fig:muons2d}
\end{figure}

Note that the scalars' maximum angular acceptance for NA62 is about $18 \, \text{mrad}$, while it is about $64 \, \text{mrad}$ for SHiP, that is, about $3.5$ times larger. Moreover, since we assume that the scalars approximately travel along the direction of their parent muons, the given muon angular spectra provide a qualitative forecast of the overall sensitivities of the two experiments.
As for the total momentum spectra in Fig.~\ref{fig:muonsmomentum}, the effective angular distribution for the experiment SHiP is more strongly affected by the IWW energy cuts than in the case of NA62. For both experiments, the IWW effect is bigger for bigger angles. To understand the relative difference of the angular spectra before and after the IWW energy cut, we consider the mutual dependence between the momentum and angular distributions analysed so far. To do so, we show the two-dimensional effective muon distributions in the plane $(P_{\mu}, \, \theta)$ in Fig.~\ref{fig:muons2d} for the experiments NA62 (left) and SHiP (right) for a fixed $m_S = 0.10 \, \text{GeV}$.

For both experiments, the low-momentum vertical strip of the two-dimensional distribution after geometrical acceptance is largest at high values of $\theta$. Thus, the IWW energy cut will generally have a stronger effect for bigger angles. 
Besides, the high-$\theta$ horizontal strip of the distribution for SHiP is more densely concentrated at the low-momentum end than it is for NA62, whose fixed-angle spectrum is more widely spread in momentum. Thus, the SHiP effective spectrum undergoes a more substantial loss due to the IWW constraint at each value of $\theta$ than the NA62 spectrum.

\section{Experimental Setups and Signatures}\label{sec:exp_acceptance}

The simplified experimental setups of NA62 and SHiP used in our toy MCs are described here, with particular attention given to the acceptance signatures of the exotic scalar's daughter particles, which result from a dedicated
detector modelling. A universal schematic diagram is shown in Fig.~\ref{fig:sketch} in Section~\ref{sec:intro}.

\subsection{NA62 Experiment run as Proton Beam Dump}

The primary goal of the NA62 experiment~\cite{NA62:2017rwk} is to make a precise measurement of the branching ratio $\mathcal{B}$($K^+ \rightarrow \pi^+ \nu \bar{\nu}$). However, the experiment is also sensitive to a variety of BSM scenarios, which appear either from the kaon decays or when NA62 is run as a beam dump~\cite{PBCNA62Massri}.
We model NA62 in a toy MC as follows. 
The beam-defining collimator for ``regular" data-taking is used to dump the beam. This collimator, which is called TAX and $L_\text{tg}=3.2~\text{m}$ long, is the source of the secondary muons that radiate the scalars. The dump material is Copper. We consider a total of $10^{19}$ protons on target (POT), as proposed to be collected between the CERN Long Shutdowns LS3 and LS4~\cite{PBCNA62}.
After a distance of $L_\text{sp}=75.8~\text{m}$ from the end of the TAX, the fiducial decay region starts, and it is $L_\text{dec}=81.0~\text{m}$ long. The scalars must decay within this region to produce potentially detectable daughter particles.
Four STRAW spectrometer chambers are placed at the end of the decay volume and are followed by a Liquid Krypton Calorimeter (LKr). The detector, composed of the spectrometer and the calorimeter, among others, has a total length of $L_\text{det}=65.2~\text{m}$ and an effective transverse acceptance area of about $2 \times 2~\text{m}^2$.
Successful tracks are required to hit the detector components at a minimum distance of $5~\text{cm}$ away from the walls of the vessel.

For the detection of the charged final states $\mu^{\pm}, \, e^{\pm}$, we impose the acceptance of both tracks in the first and last STRAW chambers and in the LKr with a minimum individual track energy of $5~\text{GeV}$. 
For the detection of the neutral final state $\gamma$, we require both photons from the scalar decay to be resolved in the LKr at a minimum mutual distance of $10~\text{cm}$. In addition, the photons need to be more than $15~\text{cm}$ away from the LKr central hole, through which the beam-pipe passes, with a minimum individual energy of $1~\text{GeV}$ and a minimum combined energy of $3~\text{GeV}$. 
Following the current knowledge based on about $10^{17}$ POT, available in the 2021 data~\cite{PBCNA62Massri}, we assume no background limitations.

\subsection{SHiP Proton Beam-Dump Experiment}

The proposed SHiP experiment~\cite{SHIP:2021tpn} specifically aims at searching for a large number of hidden new-physics states, among which weakly-interacting long-lived dark sector mediators.
The beam dump is $L_\text{tg}=11.2~\text{m}$ long and proposed to be made of Molybdenum. We assume $2 \times 10^{20}$ POT.
After a distance of $L_\text{sp}=42.0~\text{m}$ from the end of the target, the fiducial decay region starts, and it is $L_\text{dec}=50.5~\text{m}$ long. Again, the scalars must decay within this region to produce potentially detectable daughter particles.
Four STRAW spectrometer chambers are placed at the end of the decay volume and are followed by a calorimeter. The detector, composed of the spectrometer and the calorimeter, among else, has a total length of about $L_\text{det}=10.3~\text{m}$. 
Again, successful tracks are required to hit the detector components at a minimum distance of $5~\text{cm}$ away from the walls of the vessel.

For the detection of the charged final states $\mu^{\pm}, \, e^{\pm}$, we follow the guidance provided in Ref.~\cite{SHiP:2020vbd}. We require the two tracks to hit all spectrometer chambers and the calorimeter with a minimum energy of $5~\text{GeV}$ each.
The detection of the neutral final state $\gamma$ in the SHiP calorimeter is modelled following Ref.~\cite{Ahdida:2654870}. We require both photons produced by a scalar decay to hit the calorimeter within an effective elliptical acceptance area of about $5 \times 10 ~\text{m}^2$. Moreover, both photons should have an individual energy of at least $1~\text{GeV}$, a combined energy exceeding $3~\text{GeV}$, and they should be at least $10~\text{cm}$ apart. We highlight that the proposed SHiP calorimeter has the potential to reconstruct the photon direction, which in turn allows for the reconstruction of the scalar mass.
We assume no background limitations as it is inherently required by the SHiP design.

\section{Sensitivity Projections}\label{sec:results}

\begin{figure}[htb!]
\centering
 \includegraphics[width=0.85\textwidth]{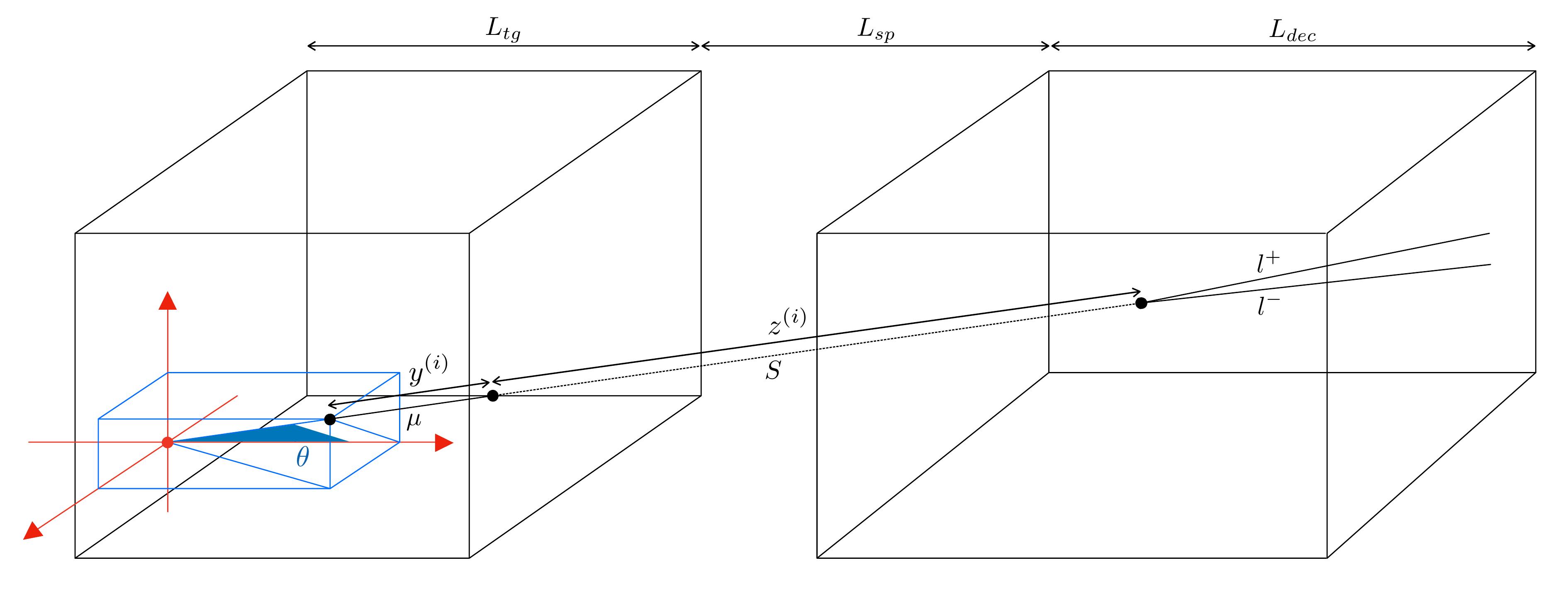}
 \caption{Schematic illustration of the experimental geometry up to scalar decay. $\theta$ is the angle between the muon trajectory and the proton beam axis. $y^{(i)}$ is the muon penetration length along its trajectory. $z^{(i)}$ is the length travelled by the scalar from production to decay along its trajectory. Lengths are not to scale.}
\label{fig:geom_sketch}
\end{figure}

In contrast to the muon beam-dump setup~\cite{Chen:2017awl}, the secondary muons generated by protons scattering on a thick target in a proton beam-dump experiment possess a whole spectrum of initial positions and momenta. To take these into account, the projected number of detected exotic signal events is computed separately for each muon from the input spectrum described in Section~\ref{sec:muons}, and the individual contributions are then summed to give the overall sensitivity projection in the parameter space $(m_S, \, g_{\mu})$ of the exotic scalar. 
We illustrate the basic geometrical quantities of interest in the simplified sketch in Fig.~\ref{fig:geom_sketch}. 

Let $N_{\mu}$ be the number of muons in the spectrum. For $i = 1, \ldots, N_{\mu}$, we denote by $E_{\mu}^{(i)}$ the energy of the $i$-th muon at a given penetration depth $y^{(i)}$ along its trajectory. See Fig.~\ref{fig:geom_sketch} for a schematic representation of $y^{(i)}$. As the muon penetrates the target, its energy $E_{\mu}^{(i)}$, as a function of $y^{(i)}$, decreases from the initial value $E_{\mu, \text{init}}^{(i)}$ at the muon production point inside the target to the minimum value $E_{\mu, \text{min}}^{(i)}$ at the end of the target. We disregard muons that are produced outside the target.
At each step in $E_{\mu}^{(i)}$, the differential cross-section for scalar production via muon bremsstrahlung $\frac{d\sigma}{dx}$ is given by the IWW approximated expression in Eq.~\eqref{eq:diffsigma} in Section~\ref{sec:cross_section}. Here, $x=E_{S}/E_{\mu}^{(i)}$.

Despite the fact that the propagating muon will have the highest energy at its creation point, the exotic scalar can, in principle, be produced at any point along the muon penetration path inside the target. The scalar production point differs from the parent muon production point by the penetration length $y^{(i)}$, which can be expressed as a function of $E_{\mu}^{(i)}$. More specifically, in the energy range that we are interested in, going from a few GeV to a few 100 GeV, ionization dominates the other mechanisms through which muons can lose their energy. In this regime, the stopping power $dE_{\mu}/dy$ is approximately constant with respect to the muon momentum~\cite{Groom:2001kq}, and we denote by $\langle dE_{\mu}/dy \rangle$ the average muon energy loss per unit of penetration length due to ionization. It follows that $y^{(i)}$ and $E_{\mu}^{(i)}$ are simply related via
\beq{ \label{eq:penetration}
    y^{(i)} \simeq \frac{E_{\mu, \text{init}}^{(i)} - E_{\mu}^{(i)}}{\langle dE_{\mu}/dy \rangle} \, .
}\eeq

We assume the scalar to be collinear to the parent muon, and we denote by $z^{(i)}$ the distance it travels along its trajectory before decay. See Fig.~\ref{fig:geom_sketch} for a schematic representation of $z^{(i)}$. The acceptance range for the scalar decay is set by the geometry of the fiducial decay volume of the specific experiment being considered, and the limiting values $z_{\text{min}}^{(i)}$ and $z_{\text{max}}^{(i)}$ are the extrapolated lengths of the straight paths which go from the scalar production point in the target to their closest and furthest intersections with the decay volume of the experiment. The geometries of NA62 and SHiP are described in Section~\ref{sec:exp_acceptance}. We disregard scalars that are not projected to intersect the decay volume.

After radiating from the muon, at each step in $z^{(i)}$, the scalar decay probability density function per unit length $P_S(z)$ is given by
\beq{ \label{eq:Ps}
P_S(z) = \frac{\text{e}^{-z/L_S}}{L_S} \, , 
}\eeq
where $L_S$ is the scalar decay length, as described in Section~\ref{sec:theory} alongside the description of the contributing decay channels. Again, at each step in $z^{(i)}$, we denote by $P_d(z)$ the experimental acceptance probability of the scalar's daughter particles conditional to their production at the given point. Specifically, we require the scalar's daughters to obey the conditions described in Section~\ref{sec:exp_acceptance} for the experiments NA62 and SHiP, which are encoded in $P_d(z)$. We assume here full reconstruction efficiency after the imposed cuts on the signal. 
Note that, in the case that all decay products are accepted, \textit{i.e.} $P_d(z)=1$, the scalar decay probability density $P_S(z)$ gives a total probability of scalar decay within the geometric acceptance range from $z_{\text{min}}^{(i)}$ to $z_{\text{max}}^{(i)}$ of 
\beq{
\int_{z_{\text{min}}^{(i)}}^{z_{\text{max}}^{(i)}} dz \, P_S(z) = \text{e}^{-z_{\text{min}}^{(i)}/L_S} - \text{e}^{-z_{\text{max}}^{(i)}/L_S} \, .
}\eeq
Finally, the total number of detected exotic signals for a given choice of parameters $(m_S, \, g_{\mu})$ is produced by the following composition of the scalar production differential cross-section and the convolution of the scalar decay probability density with the daughters' experimental acceptance efficiency
\beq{ \label{eq:Ns_tot}
N_S = \frac{n_{\rho}}{\langle dE_{\mu}/dy \rangle} {\sum_{i=1}}^{N_{\mu}}} {\int_{E_{\mu, \text{min}}^{(i)}}^{E_{\mu, \text{init}}^{(i)}} dE_{\mu} \; {\int_{0}^{1}} dx \, \frac{d \sigma}{dx} \; {\int_{z_{\text{min}}^{(i)}}^{z_{\text{max}}^{(i)}}} dz \, P_S(z) \cdot 
P_d(z) \, ,
}\eeq
where $n_{\rho}$ is the number density of the target nuclei.

\begin{figure}[htb!]
\centering
\subfloat[Leptophilic Model]{\includegraphics[width=0.49\textwidth]{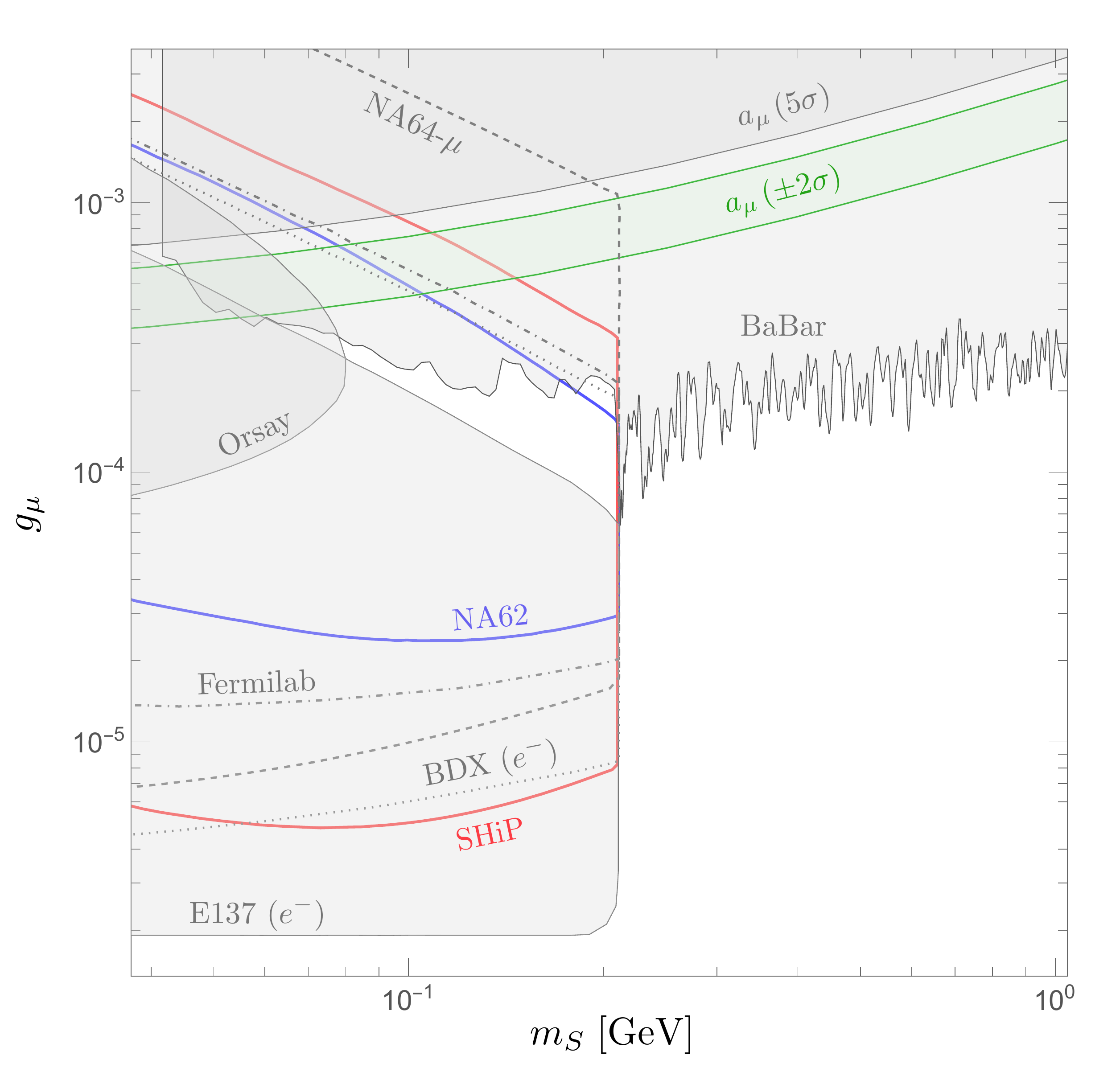}}
\quad
\subfloat[Muonphilic Model]{\includegraphics[width=0.49\textwidth]{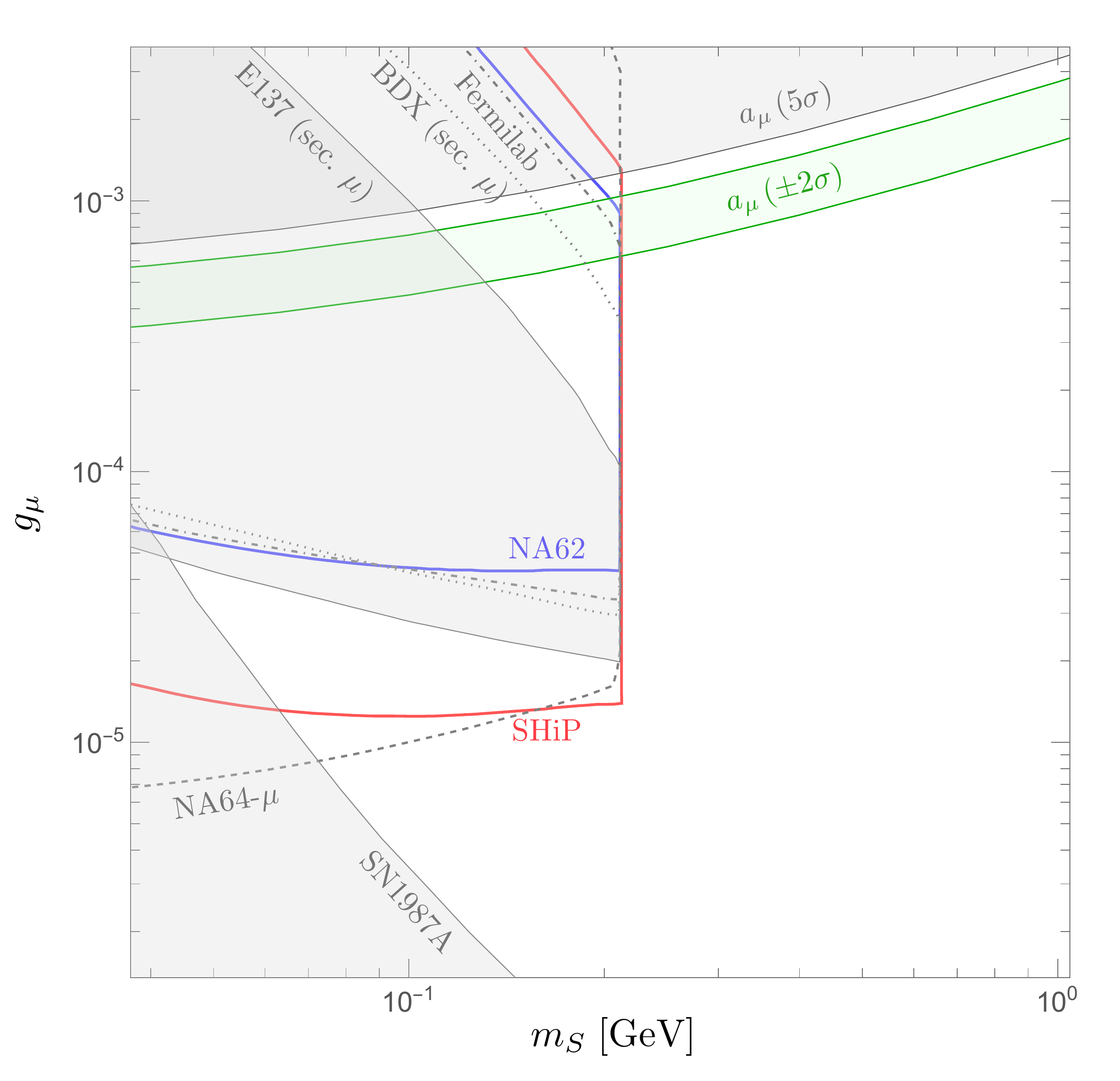}}
 \caption{Projected and experimental constraints in the $(m_S, \, g_{\mu})$-plane for (a) the leptophilic model, where $g_\ell \propto m_\ell$, and (b) the muonphilic model, where $g_\mu \propto m_\mu, \, g_e=g_\tau=0$. Our projections for NA62 and SHiP are cut at the di-muon mass $m_S = 2 m_\mu$ and are shown as solid blue and red lines, respectively. The gray shaded regions denote the strongest existing bounds on the new scalar mediator by various beam-dump, collider, and astrophysical probes. The gray non-continuous lines represent a selection of current prospects, which are included for comparison. See the text for details. We also show the $5\sigma$-excluded region (shaded gray) and $2\sigma$-favored region (shaded green) for the anomalous $(g-2)_\mu$ measurement by Fermilab and Brookhaven~\cite{PhysRevLett.126.141801}.}
\label{fig:projections}
\end{figure}

Along with the many geometric and experimental cuts outlined above, the computational implementation of Eq.~\eqref{eq:Ns_tot} is helped by the double cut on the muon energy that is required in order to apply the IWW approximation for our predictions, as described in Section~\ref{sec:cross_section} and furthermore analysed in Section~\ref{sec:muons}. 
The resulting conservative estimates of the number of signal events $N_S$ for exotic scalar production and detection at the experiment NA62 run in beam-dump mode and at the proposed SHiP beam-dump facility, for both the leptophilic and muonphilic effective models of scalar couplings, that have been produced with the present work, are translated into projected sensitivity constraints in the parameter plane $(m_S, \, g_{\mu})$. After cutting at the di-muon mass $m_S = 2 m_\mu$, the corresponding exclusion contours at $95\%$ confidence level are presented in Fig.~\ref{fig:projections} alongside current prospects and constraints from other experiments in the current literature.

In Fig.~\ref{fig:projections}, the strongest existing constraints are shown as gray shaded regions. For the leptophilic model, they come from the beam-dump experiments at Orsay~\cite{DAVIER1989150} and E137~\cite{PhysRevD.38.3375}, as well as collider searches by the BaBar collaboration~\cite{BaBar:2020jma}. For the muonphilic model, they arise from the E137 search and from experimental limits set on the anomalous cooling of SN1987A due to low-mass muonphilic scalars~\cite{Croon:2020lrf}\footnote{More recent work on supernova bounds for muonphilic scalars can be found in Ref.~\cite{PhysRevLett.128.221103}.}. We show the $2\sigma$-favored region for the anomalous $(g-2)_\mu$ measurement by Fermilab and Brookhaven~\cite{PhysRevLett.126.141801} in solid shaded green and the corresponding $5\sigma$-excluded region in solid shaded gray.
For comparison, we include the projections for the proposed BDX search~\cite{Marsicano:2018vin} (gray dotted line), which considers scalar bremsstrahlung from secondary muons coming from an electron beam, and the projections for the scalar searches using a muon beam on a dump at NA64-$\mu$ (gray dashed line) and Fermilab (gray dash-dotted line)~\cite{Chen:2017awl, Chen:2018vkr, Kahn:2018cqs}.  
We remark that our conservative projections for both NA62 (solid blue line) and SHiP (solid red line) span uncharted parameter space for both models. In the muonphilic case, they intersect a part of the $2\sigma$-favored region of parameter space which has not been rejected by existing experimental bounds, thus accessing a new-physics sector potentially responsible for the $(g-2)_\mu$ anomaly, as discussed in Section~\ref{sec:intro}. 

Note that the upper profile of the beam-dump prospects, corresponding to large coupling $g_\mu$, is mostly determined by the experimental geometry, while the lower contour, corresponding to very small coupling, has a different slope for primary- and secondary-muons set-ups.
The former typically have a rather monochromatic muon spectrum, while the latter possess a wide range of angular and energetic muon components. For a given mass $m_S$, the scalars radiated from secondary muons are generally less boosted, giving rise to the difference in the slope of the lower shape.
Moreover, we stop our projections at the di-muon threshold $m_S = 2 m_\mu$ for ease of comparison with previous prospects. In principle, both NA62 and SHiP will have sensitivity to muon final states, as described in Section~\ref{sec:exp_acceptance}. Expanding our analysis to include muon final states would allow us to extend our reach beyond the di-muon mass, thus probing further unexplored parameter space. However, the IWW approximation for the computation of the scalar production cross-section is less robust in this region, and the evaluation of the sensitivity beyond the di-muon threshold requires additional computational resources. We leave this extension of our results for future work. 

\section{Conclusions}\label{sec:conclusions}

In this paper, we have explored the potential of using the secondary muons of a proton beam-dump experiment as a practicable and competing alternative signal production source in the search for muon-specific extensions of the SM. BSM models of long-lived and low-mass scalar particles that are primarily or exclusively coupled to muons are prompted on experimental and theoretical grounds, among which is the standing discrepancy between the SM prediction and the experimental observation of the anomalous magnetic moment of the muon.
In contrast to searches using a dedicated muon beam, we have proposed to exploit the shower of secondary muons that are created by the reactions of protons on target in a proton beam dump, as sketched in Fig.~\ref{fig:sketch} in Section~\ref{sec:intro}.

After outlining the two simplified models of exotic scalar couplings to leptons that we have denoted as leptophilic and muonphilic models in Section~\ref{sec:theory}, and which can potentially alleviate, or even resolve, the $(g-2)_\mu$ anomaly, we have described the muon bremsstrahlung process which dominates the scalar production mechanism. We have introduced and justified, within its limits of applicability, the IWW approximation scheme for the computation of the scalar production cross-section in Section~\ref{sec:cross_section}, which is used to obtain the results presented in this work. For our purposes, and given the input energy distribution of the secondary muons and the geometrical acceptance of the experiments NA62 and SHiP under analysis, as closely inspected in Sections~\ref{sec:muons} and~\ref{sec:exp_acceptance}, the IWW approximation provides a good estimate of the exact solution. However, we expect that a more rigorous study incorporating the full scalar production cross-section via complete MC simulations will improve upon our conservative projections, especially above the di-muon mass threshold. We leave such an improvement to future work.
For the thorough modelling of the original muon flux from the proton beam dump reviewed in Section~\ref{sec:muons}, we have relied on the state-of-the-art MC simulation provided in Ref.~\cite{stealex_2022_6352043}, which efficiently produces high statistics muon samples.
In Section~\ref{sec:results}, we have computed the sensitivity reach of the currently running NA62 experiment in beam-dump mode as well as the proposed SHiP beam dump.
We have shown in Fig.~\ref{fig:projections} that proton beam dumps can be competitive with primary muon-beam experiments in the probe for dark muon-specific scalars. Indeed, our conservative projections for both NA62 and SHiP cover unexplored parameter space for both the leptophilic and the muonphilic models. Moreover, for scalars with exclusive coupling to muons, our prospects intersect a part of the $2\sigma$-favored region of parameter space which has not been rejected by existing experimental bounds, roughly centered around $m_S \sim 160 \, \text{MeV}$ and $g_\mu \sim 7 \times 10^{-4}$.
Our result enlarges the class of BSM models that can be effectively probed by proton beam dump experiments and, in particular, gives access to a new-physics sector potentially responsible for the $(g-2)_\mu$ anomaly.

We remark that a major additional challenge for the proton beam-dump setting, as compared to muon facilities, is a reliable modelling of the secondary muon spectrum. Addressing this challenge entails the need for accurate computational tools that are validated in experimental set-ups. Because NA62, for example, has already collected a significant amount of beam-dump data in 2021 and is progressing swiftly, the work presented in this paper is particularly timely~\cite{DAlessandro:2021gmw}.

Finally, we note that it could be interesting to interface the scalar production process from secondary muons presented here with the ALPINIST framework~\cite{Jerhot:2022chi}, where the model-dependent and the model-independent components of the sensitivity evaluation for pseudo-scalar (ALP) production and decay are separated. This allows a somewhat smooth integration of new models with non-trivial coupling structures, such as the ones studied in this paper, and detector geometries with optimised computational effort.
The currently missing ingredient in order to realise such an interface is the secondary muon spectrum coming from proton beams with lower energies, although this could in principle be obtained. We leave this venture to future work.

\section*{Acknowledgments}

We acknowledge the contributions of Samantha I. Davis and Marcel Rosenthal to a previous version of the project.
We are grateful to Stefan Ghinescu for his effort and help with the production of the muon samples.
We thank Joerg Jaeckel for discussions that led to the project idea and Sam McDermott and Brian Shuve for general discussions. 
We also thank Yiming Zhong for providing the {\tt MadGraph} files needed to simulate the cross-section and for his patience in troubleshooting. 
This work has been supported by the Swiss National Centre of Competence in Research SwissMAP (NCCR 51NF40-141869 The Mathematics of Physics), the European Research Council under Grant No. ERC-2018-StG-802836
(AxScale project), and the National Science Foundation under Grant No. NSF CAREER-1944826-PHY.

\bibliographystyle{utphys}
\bibliography{main.bib}

\end{document}